\documentclass[fleqn,10pt]{wlscirep}
\title{Effect of FePd alloy composition on the dynamics of artificial spin ice}

\author[1,2,$\dagger$]{Sophie~A.~Morley}
\author[3]{Susan~T.~Riley}
\author[4,5]{Jose-Maria~Porro}
\author[3]{Mark~C.~Rosamond}
\author[3]{Edmund~H.~Linfield}
\author[3]{John~E.~Cunningham}
\author[4]{Sean~Langridge}
\author[1,*]{Christopher~H.~Marrows}
\affil[1]{School of Physics and Astronomy, University of Leeds, Leeds LS2 9JT, UK}
\affil[2]{Department of Physics, University of California, Santa Cruz, California, 95064, USA}
\affil[3]{School of Electronic and Electrical Engineering, University of Leeds, Leeds LS2 9JT, UK}
\affil[4]{ISIS Neutron and Muon Source, STFC, Rutherford Appleton Laboratory, Harwell Science and Innovation Campus, Didcot OX11 0QX, UK}
\affil[5]{BCMaterials, Basque Center for Materials, Applications and Nanostructures, 48160 Derio, Spain}

\affil[$\dagger$]{samorley@ucsc.edu}
\affil[*]{c.h.marrows@leeds.ac.uk}

\begin{abstract}

Artificial spin ices (ASI) are arrays of single domain nano-magnetic islands, arranged in geometries that give rise to frustrated magnetostatic interactions. It is possible to reach their ground state via thermal annealing. We have made square ASI using different FePd alloys to vary the magnetization via co-sputtering. From a polarized state the samples were incrementally heated and we measured the vertex population as a function of temperature using magnetic force microscopy. For the higher magnetization FePd sample, we report an onset of dynamics at $T = 493$~ K, with a rapid collapse into $> 90$~\% ground state vertices. In contrast, the low magnetization sample started to fluctuate at lower temperatures, $T = 393$~K and over a wider temperature range but only reached a maximum of 25$\%$ of ground state vertices. These results indicate that the interaction strength, dynamic temperature range and pathways can be finely tuned using a simple co-sputtering process. In addition we have compared our experimental values of the blocking temperature to those predicted using the simple N\'{e}el-Brown two-state model and find a large discrepancy which we attribute to activation volumes much smaller than the island volume. 

\end{abstract}
\begin{document}

\flushbottom
\maketitle
\thispagestyle{empty}

\section*{Introduction}
Over the past decade there have been many reports on a new type of nanomagnetic system which presents a novel way to design frustration using a top-down approach, the artificial spin ices (ASI). These systems can either mimic more complex and atomic 3D systems such as the spin ice materials \cite{Harris1997,realspin}, after which they were first named, or more recently be used to create completely new emergent behavior via sample design \cite{Gilbert2016, Gilbert2014, Nisoli2013}. ASI maintain their popularity as an exciting experimental field due to the immense tunability of the system parameters such as dimensions \cite{Morley2015,farhansquare}, material \cite{melting,Porro2013,Drisko2015}, vertex coordination \cite{Gilbert2014,Gilbert2016}, substrate and geometry \cite{Zhang2013, Farhan2016, Gliga2017, Shi2017}, which present new and uncharted paths to explore.

In order to access the interesting physics of these systems it is necessary to reduce the energy barrier between the two uniaxial stable states of an island to be somewhere close to the thermal energy available, so that fluctuations are possible and real thermodynamic processes can take place. This energy barrier is defined as $KV = \frac{1}{2} \mu_{0} \Delta N_{D} M_{\mathrm{S}}^{2} V$. Here $K$ is the (shape) anisotropy constant, $V$ is the island volume, $\Delta N_{D}$ is the difference in demagnetizing factors between the easy and next least hard directions, and $M_{\mathrm{S}}$ is the saturation magnetization of the material from which the island is formed. Most early work used permalloy (Ni$_{80}$Fe$_{20}$) as the island material, whose Curie temperature is relatively high; $T_{\mathrm{C}} = 873$~K. This presented a problem, as it required making either very small nano-islands (to lower $V$), which would be difficult to microscopically inspect or to use exceedingly high temperatures (to sufficiently lower $M_{\mathrm{S}}$). These samples were usually grown on silicon substrates which meant the relatively high temperatures required led to problems with inter-diffusion and the destruction of the magnetism of the islands. 

However, a long-range ground state was discovered in an as-grown sample where the islands had a small enough volume during growth that they were able to form large domains of ground state before the final sample thickness was reached and the order was frozen-in \cite{morgan}. Other early examples of circumventing the anneal problem included using a NiFe alloy with a lower magnetization and $T_{\mathrm{C}}$ \cite{Porro2013} or using a protective SiN buffer between the substrate and the sample \cite{Zhang2013}. Exploiting the fact that the islands would be thermally active when having smaller volumes, Farhan et al. \cite{farhansquare, hypercube} made a wedge sample in order to find the exact thickness where dynamics occurred around room temperature. They found for relatively large lateral islands that this thickness was $t \approx 3$~nm. For islands with relatively large lateral dimensions but very thin it is possible to use advanced magnetic microscopy techniques at synchrotrons such as photo-emission electron microscopy (PEEM) \cite{buildingblocks, farhansquare, Gilbert2016, Kapaklis2014} in order to track real-time thermal magnetic processes. An alternative solution was to choose a material different from Py, such as Fe delta layers in Pd \cite{buildingblocks, Kapaklis2014}. However, this material tends to have a low magnetization and this can be at the expense of interaction strength between islands. 

The current study was inspired by the work of Drisko \textit{et al.}, who showed that the Curie temperature and magnetization of an FePd alloy can be tuned in order to control the level of ordering within an ASI system, whilst maintaining strong interactions \cite{ Drisko2015}. We used a co-sputtering technique to deposit thin films of various compositions of FePd via control of relative growth rates of Fe and Pd. In order to fabricate the ASI samples we used a hard mask process and Ar ion milling. Each thermalization step was carried out in a rapid thermal annealer with a N$_{2}$ atmosphere to prevent oxidation. The sample was held for 2 minutes at the desired temperature, after which it was cooled back down to room temperature and its resultant magnetic configuration was then imaged using the magnetic force microscopy (MFM). Using this method we have been able to track the vertex population dynamics as the temperature was changed by using a combination of material manipulation and a simple method without the use of a synchrotron. By doing so we have shown it is possible to obtain detailed information on the dynamic processes by a simple alteration of the material growth. 

\begin{figure}[tb]
\centering
\includegraphics[width=17cm]{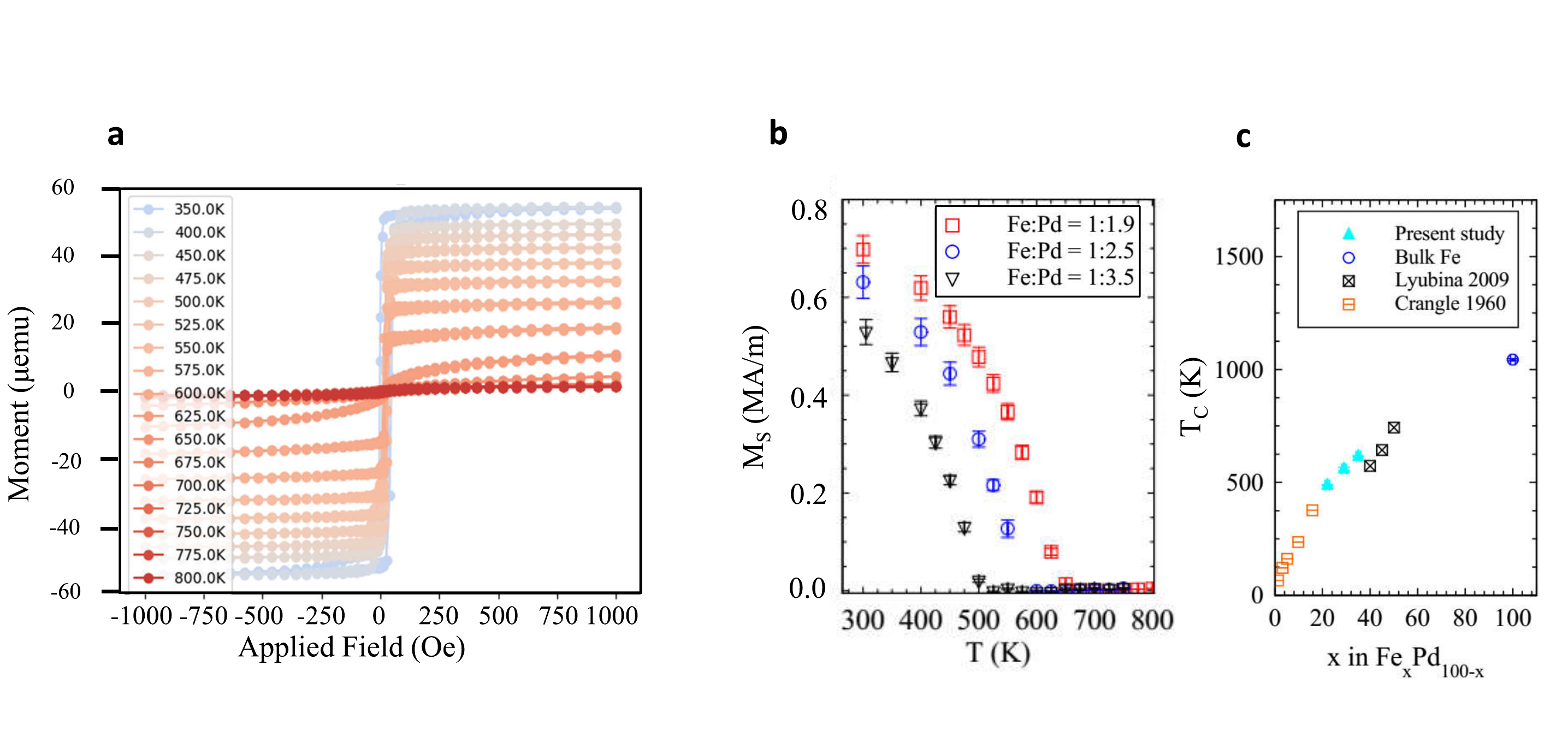}
\caption{Magnetic properties of FePd alloy films. (a)~Hysteresis loops taken at different temperatures for the Fe:Pd~=~1:1.9 thin film sample. (b)~Saturation magnetization as a function of temperature for three different compositions of FePd alloy. (c)~The extracted Curie temperatures, $T_{C}$, for the various compositions plotted with those taken from literature \cite{Crangle1960,Lyubina2009}.}
\label{fig:mag_comp}
\end{figure}

\section*{Results}

\subsection*{Thin film magnetic properties}

Thin films of different FePd compositions were grown via sputtering co-depostion. The composition of FePd can be tuned in order to control the magnetization and Curie temperature of the material \cite{Crangle1960}. In order to compare the relative magnetic strengths of the different alloys that were grown, the moment as a function of field was measured at different temperatures $T$ using a superconducting quantum interference device vibrating sample magnetometer (SQUID-VSM). The data for lowest Pd-content alloy are shown in Fig.~\ref{fig:mag_comp}a. The film thicknesses were measured using low-angle x-ray reflectivity and the areas were measured using image analysis software of photographs of the samples in order to calculate the $M_{\mathrm{S}}$ for each film. The temperature dependences of this quantity are plotted in Fig.~\ref{fig:mag_comp}b for the high, medium and low magnetization samples, labeled with their Fe:Pd ratios, 1:1.9, 1:2.5 and 1:3.5, respectively. These ratios were calculated from the measured growth rates, as described in the Methods section. The films have between 22 and 34~$\%$ Fe content.

As is expected, the sample with the lower Fe content (1:3.5 ratio) has a smaller magnetization over the entire measurement range and its magnetization at room temperature is $M_{\mathrm{S}}$(300~K)~=~5.29~MAm$^{-1}$. For the highest magnetization sample (1:1.9 ratio), $M_{\mathrm{S}}$(300~K)~=~6.98~MAm$^{-1}$, which equates to a 38$\%$ increase in magnetization for an increase from $x$=22 to 34 in the Fe$_{x}$Pd$_{100-x}$ alloy. These agree well with those values found in the literature for similar compositions \cite{Crangle1960,Myagkov2012}.

We were able to fit the well-known $M(T) \propto (T_{C} - T)^{1/2}$ law (derived from a simple Landau theory approach) to the data in order to extract the $T_{\mathrm{C}}$ of the films. These extracted values from the present study (solid triangles) are plotted along with bulk values taken from the literature \cite{Crangle1960,Lyubina2009}. There is good agreement between our thin film values and those of the bulk, indicating the relative growth rates give a reliable estimate of the Fe:Pd ratio. 

\begin{figure}[tb]
\centering
\includegraphics[width=15cm]{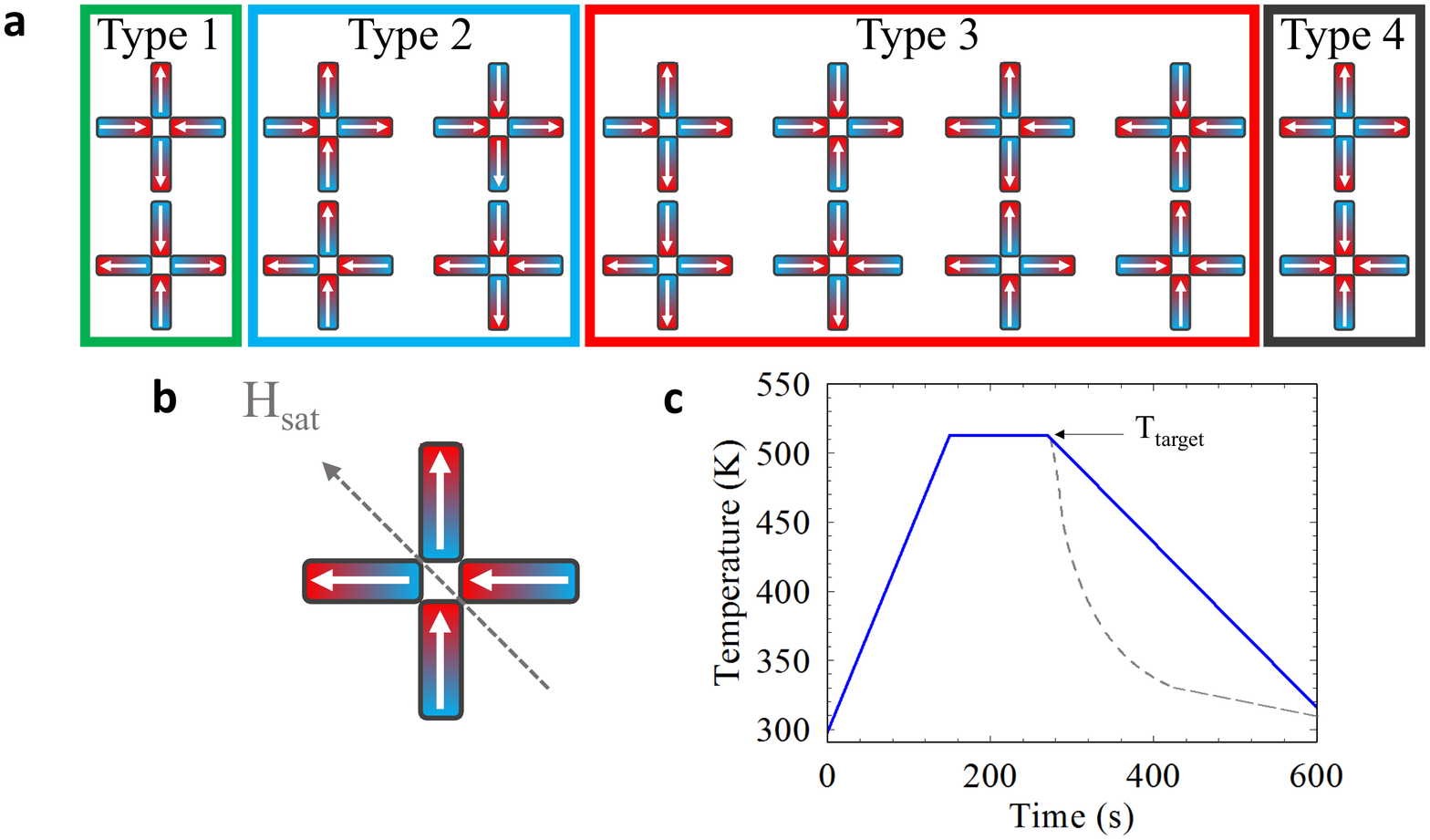}
\caption{Artificial spin ice configurations and annealing. (a) The 16 different possible vertex configurations grouped by increasing energy from Type 1 to 4. (b)~The direction of the applied field in order to prepare the initial T$_{2}$ saturated state. (c)~An example of an anneal process carried out where the target temperature was $T$ = 513~K. The heating was done over a 120~s period, the sample was held at the target temperature for 180~s and the cooling duration was 3~minutes, where the initial cooling took place more rapidly, indicated by the dotted line.}
\label{fig:anneal_protocol}
\end{figure}

\subsection*{Artificial spin ice measurements}

We used the highest and lowest magnetization sample to pattern square artificial spin ice. We fabricated islands with dimensions 80~$\times$~240~nm$^{2}$ and thicknesses $\sim$~30~nm (see Methods) and lattice spacings of $a$~=~450, 500, and 600~nm. The lithography was carried out using a hard Ti mask and broad-beam ion milling. We patterned the square geometry where each vertex has four nanomagnets whose moments can point either in or out of the vertex. As each macrospin of the nanomagnet has two possible orientations (aligned parallel or anti-parallel to the longest side) and there are four nanomagnets which contribute to a vertex, so they can be arranged in a total of 2$^{4}$~=~16 different ways, as shown in Fig.~\ref{fig:anneal_protocol}a. The so-called `ice-rule' obeying states are those which have two macrospins which point into the vertex and two which point out. It is possible to have the ground state energy of the vertex when the two pointing in (or out) are parallel to each other, type 1 (T$_{1}$). The polarized state has a slightly higher energy owing to the two like-poles meeting on the closer perpendicular neighbor, type 2 (T$_{2}$) but still obeys the `ice-rule'. We can set all of the vertices into the same T$_{2}$-state by applying and then removing a large saturating field along the diagonal of the array (as indicated in Fig.~\ref{fig:anneal_protocol}b). The samples were then heated, using a rapid thermal annealer to a target temperature for two minutes and cooled back to room temperature over a three minute period. The target temperatures investigated were in the range $T_{\mathrm{target}}$=~353~-~513~K. An example of the ramp protocol for $T_{\mathrm{target}}$=~513~K is shown in Fig.~\ref{fig:anneal_protocol}c, with the actual attained cooling measured via a thermocouple in contact with the sample plate plotted with a dotted line. The magnetic configurations of the samples were subsequently imaged at room temperature using MFM.

\begin{figure}[tb]
\centering
\includegraphics[width=18cm]{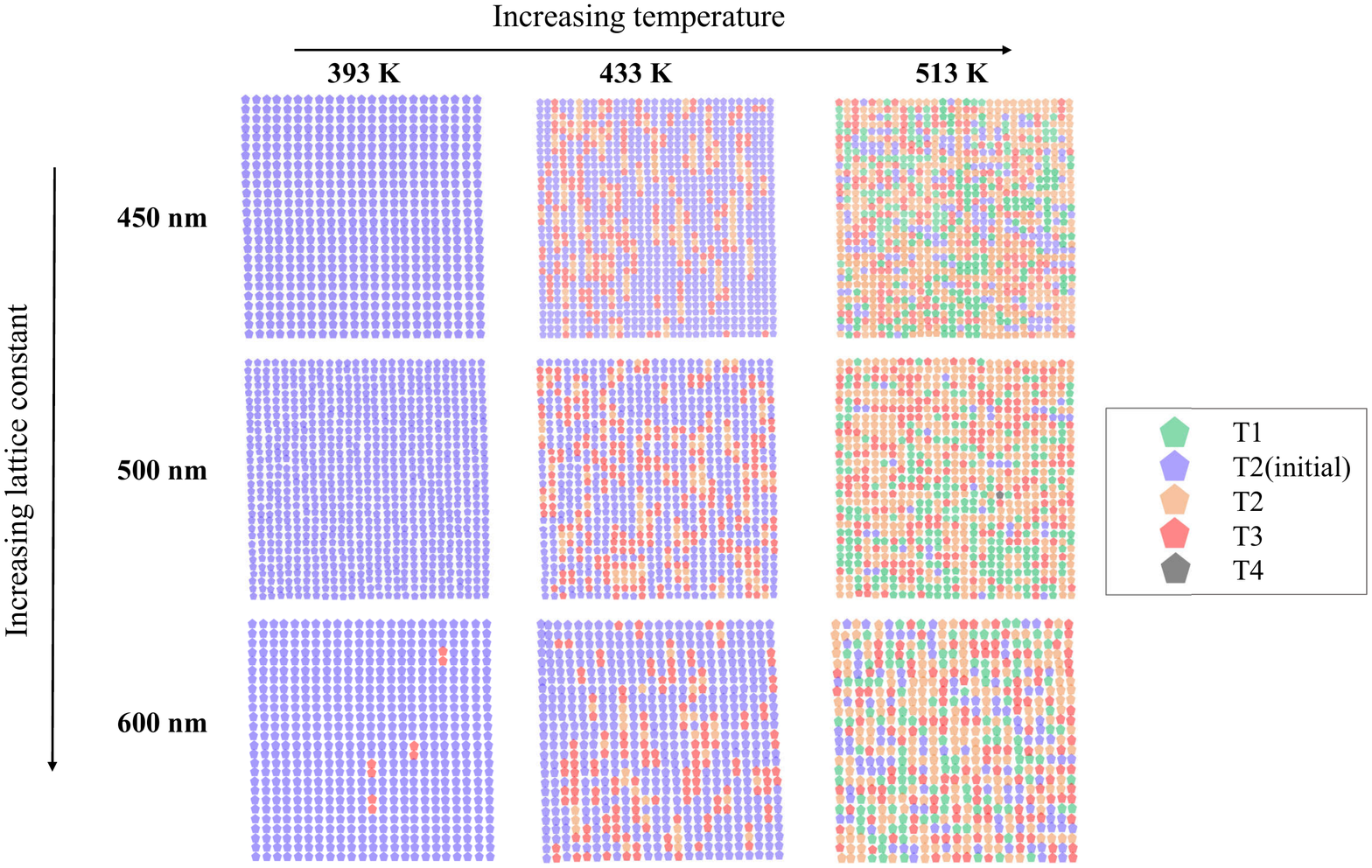}
\caption{Room temperature MFM images of the Fe:Pd = 1:3.5 arrays after annealing at various temperatures, shown for increasing lattice constant (top to bottom), each image is 15$\times$15~$\mu$m$^2$. After an anneal at $T$~=~323~K, all the islands remain unchanged and in the polarized state. As the anneal temperature was increased the least interacting sample, $a$~=~600~nm, changes first, with T$_{3}$ vertices being created at $T$~=~393~K. Samples of all lattice spacings become highly disordered after annealing at the highest temperature, $T$~=~513~K.}
\label{fig:lowmag_MFM}
\end{figure}

\begin{figure}[tb]
\centering
\includegraphics[width=18cm]{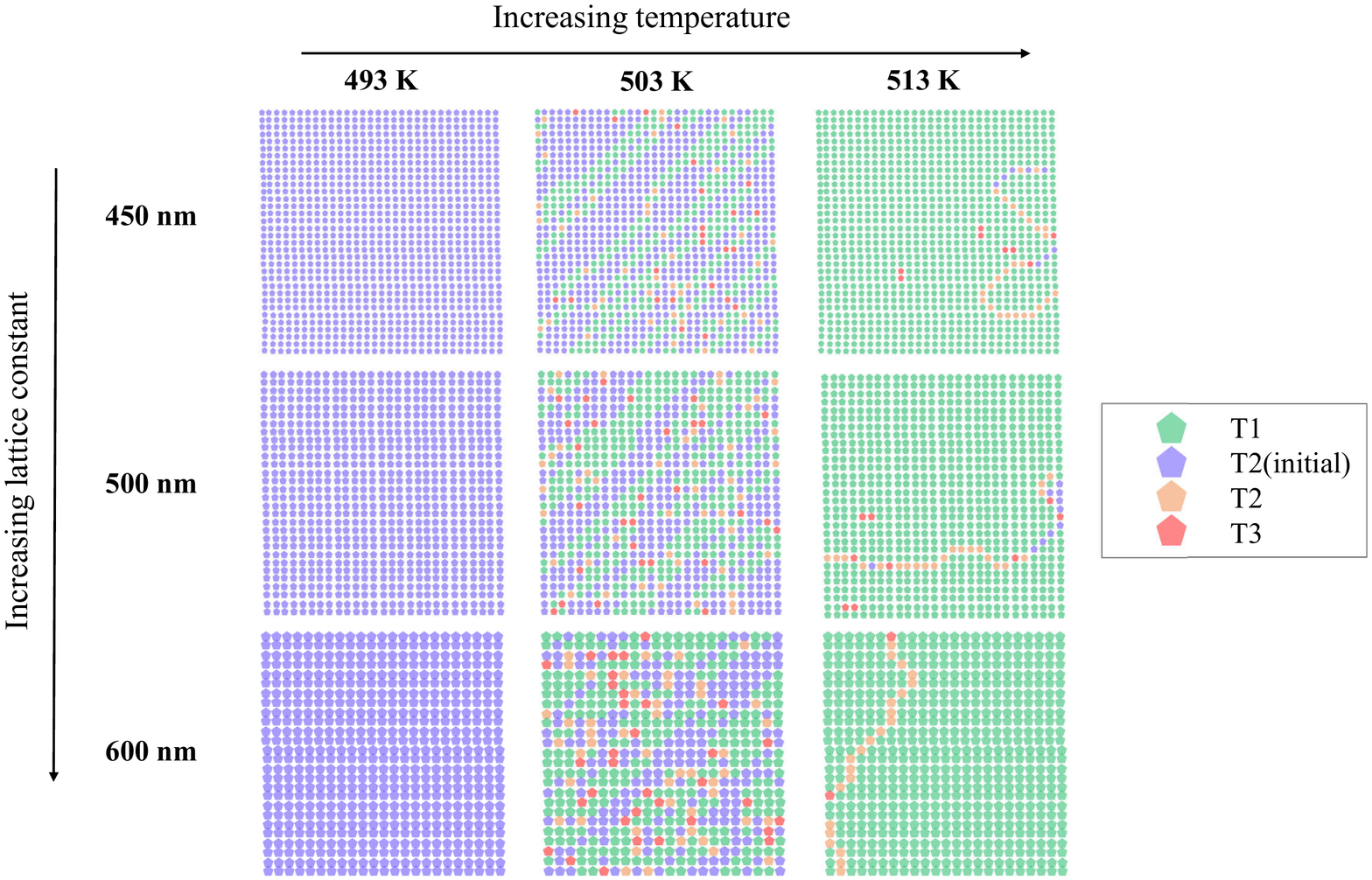}
\caption{The vertex map for the high magnetization sample (Fe:Pd = 1:1.9) as a function of anneal temperature step from left to right, $T$= 493, 503, and 513~ K; and with varied lattice constants from top to bottom, $a$ = 450, 500, and 600 nm. Each map is based on an MFM image of 15$\times$15~$\mu$m$^2$ and starts as a fully polarized T$_{2}$ background (blue dots) at the lowest temperature and thermally relaxes into almost entirely ground state T$_{1}$ vertices (green dots) within a small temperature window, $\Delta T$ = 20~K. Domain walls made up of T$_{2}$ and T$_{3}$ vertices separate ground state domains of different chirality in the final state.}
\label{fig:highmag_MFM}
\end{figure}

The resulting MFM images after each temperature step are shown for three different lattice spacings in the low magnetization sample (Fe:Pd = 1:3.5) in Fig.~\ref{fig:lowmag_MFM} (each image is 15$\times$15~$\mu$m$^2$). All samples start in the fully polarized T$_{2}$ state. We can see the anneal temperature for the onset of magnetic relaxation for the least interacting sample ($a$~=~600~nm) is $T$~=~393~K, where four pairs of emergent magnetic charges of opposite sign T$_{3}$ vertices (also referred to as emergent monopole-antimonopole pairs) have been created on top of the otherwise uniform T$_{2}$ background. For the $a$~=~450 and 500~nm, the same number appear in the next temperature step $T$~=~413K. As the temperature is increased they predominantly dissociate by vertical paths in the image which creates a string of different T$_{2}$ vertices in their wake with T$_{3}$ vertices at their ends, this is much less pronounced in the least interacting sample. On increasing the temperature further we observed a high level of disorder and an overall vertex population consisting of a mixture of mainly T$_{1}$, T$_{2}$ and T$_{3}$ vertices. 

The same experiment was carried out for the high magnetization sample with Fe:Pd = 1:1.9. The resulting evolution of the vertex populations is shown for the same three lattice spacings in Fig.~\ref{fig:highmag_MFM}. The sample was measured after anneals starting from $T$~=~353~K, however there was no change in any of the vertex states until $T$~=~503~K. This onset temperature is 110~K higher than that of the $a$~=~600~nm low magnetization sample and 90~K higher than the onset temperature of the other lattice spacings of the low magnetization sample. As can be seen from the figure a large proportion of the islands (\textgreater90~\%) reach the ground state (green dots) by the final temperature step $T$~=~513~K. The sample has a tendency to relax, particularly in the two smallest lattice spacing samples, via the dissociation of emergent monopole-antimonopole pairs similar to the lower magnetization sample. However, this time the resulting string propagation was across the diagonal and created T$_{1}$ vertices in its wake, as previously seen in the thermal relaxation of strongly interacting square ice \cite{farhansquare}. This allows large areas of ground state domain to form after the initial strings have been created.

\section*{Discussion}

First, we can compare the theoretical estimate of the blocking temperature with that which we have measured using the equation defined by N\'eel and Brown:
\begin{equation}
T_{\mathrm{B}} = \frac{KV} {\ln (t_{m}/t_{0}) k_{\mathrm{B}}},
\label{eq:N-B}
\end{equation}
where $t_{m}$ is the measurement time of the moment, which we take as the wait time of two~minutes, $t_{0}$~=~10$^{-10}$s is a characteristic relaxation time \cite{coey}, $T_{\mathrm{B}}$ is the blocking temperature of the array, $K$ is the anisotropy constant which depends on the material parameters and difference in the demagnetization factors from the shape ($K=\frac{1}{2}\mu_{0} \Delta N_{D} M_{\mathrm{S}}^{2}$) with $\Delta N_{D}$~=~0.15 \cite{osborn} and $k_{B}$ is Boltzmann’s constant. For the low magnetization sample, we estimate the blocking temperature as the temperature step where half of the vertices have changed from their initial state $\sim$~443~K. Rearranging Eq.~\ref{eq:N-B}, we can estimate the magnetization from the measured blocking temperature, $M_{\mathrm{S}}$~=~53$\pm$4~kA/m which we can compare to the measured value of the film at the same temperature, $M_{\mathrm{S}}(443K)$~=~260~kA/m. This differs by an order of magnitude which means the barrier to flip the island is much smaller than expected. Doing the same analysis for the high magnetization sample we find the $M_{\mathrm{S}}$~=~60$\pm$4~kA/m from Eq.~\ref{eq:N-B} compared to the $M_{\mathrm{S}}(503K)$~=~470~kA/m which shows an even larger discrepancy with the model. A similarly large discrepancy was recently reported in ultra-small nanomagnets \cite{Morley2017}. The reason for such a discrepancy is rarely discussed, however we propose it is due to the oversimplified treatment of the internal spin configuration of the islands. It has recently been shown using ferromagnetic resonance (FMR) studies that there is a pronounced bending of the magnetization towards the edges of the islands \cite{Gliga2015}; this canting at the ends may cause metastable states to initiate the coherent reversal at lower temperatures than that which is predicted from a simple two-state Neel-Brown model. For instance, to observe a blocking temperature which relates to the calculated $M_{\mathrm{S}}$, we would need a reduction of the volume which corresponds to $\sim$~30~nm$^{3}$ activation volume for the low $M_{\mathrm{S}}$ sample and $\sim$~20~nm$^{3}$ activation volume for the high $M_{\mathrm{S}}$ sample. Activation volumes much smaller than the island, such as these sizes, could potentially correspond to that of a crystal grain and should be taken into account when designing thermal ASI systems.

\begin{figure}[tb]
\centering
\includegraphics[width=16cm]{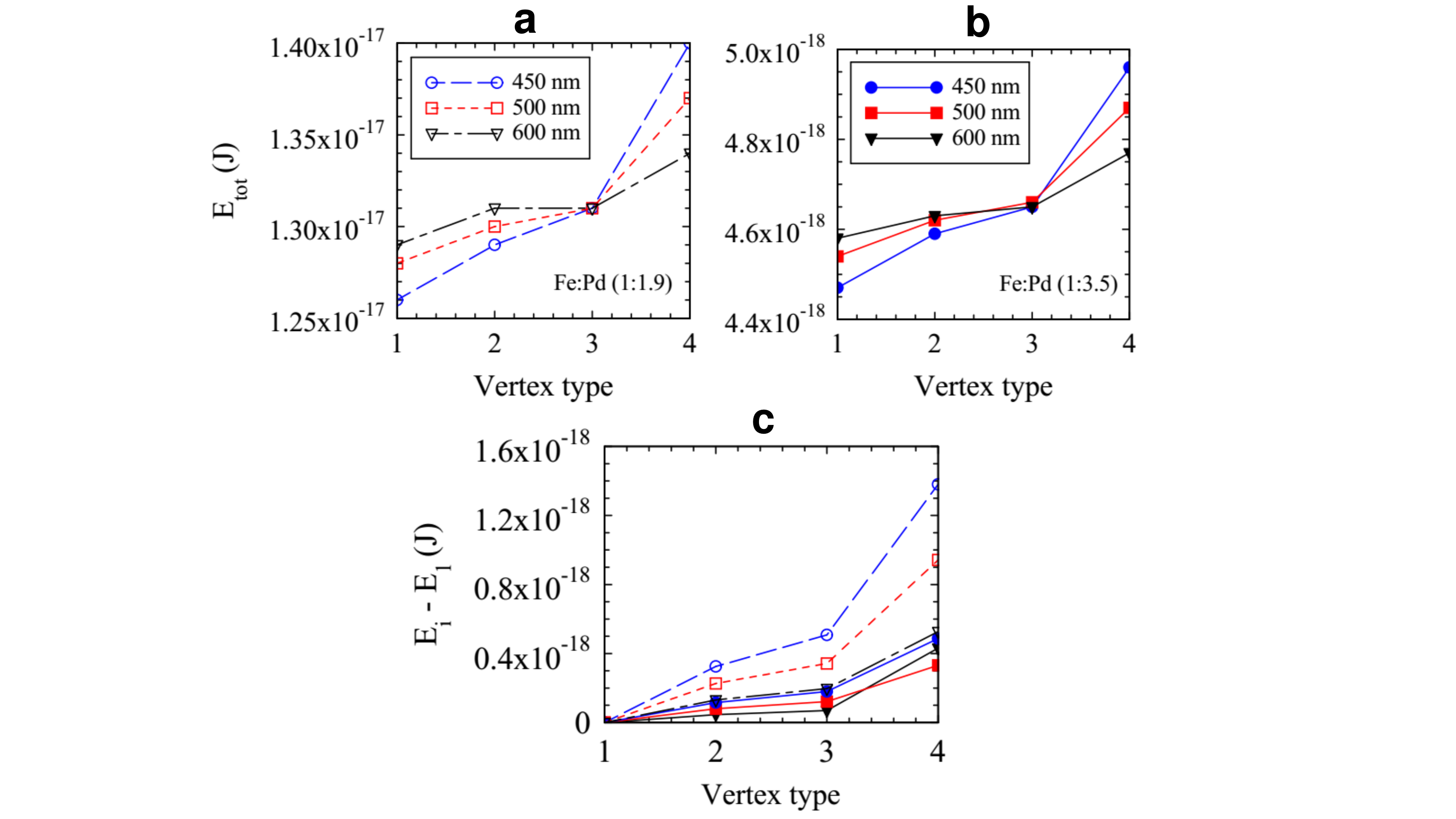}
\caption{Vertex energies calculated using OOMMF simulations. We have taken the value of $M_{\mathrm{S}}$ at the temperature where the string propagation occurs for each set of samples. These energies are plotted in (a) the high magnetization sample where  $M_{\mathrm{S}}$ was taken at 503~K and (b) for the low magnetization sample where  $M_{\mathrm{S}}$ was taken at 433~K. (c) The difference in energy between the lowest energy vertex, type 1, and the other vertex types, $E_{i} - E_{1}$ for the high magnetization sample (open symbols) and the low magnetization sample (filled symbols), for the three lattice spacings studied $a$ = 450, 500 and 600nm.}
\label{fig:vertexenergy}
\end{figure}

\begin{figure}[tb]
\centering
\includegraphics[width=14cm]{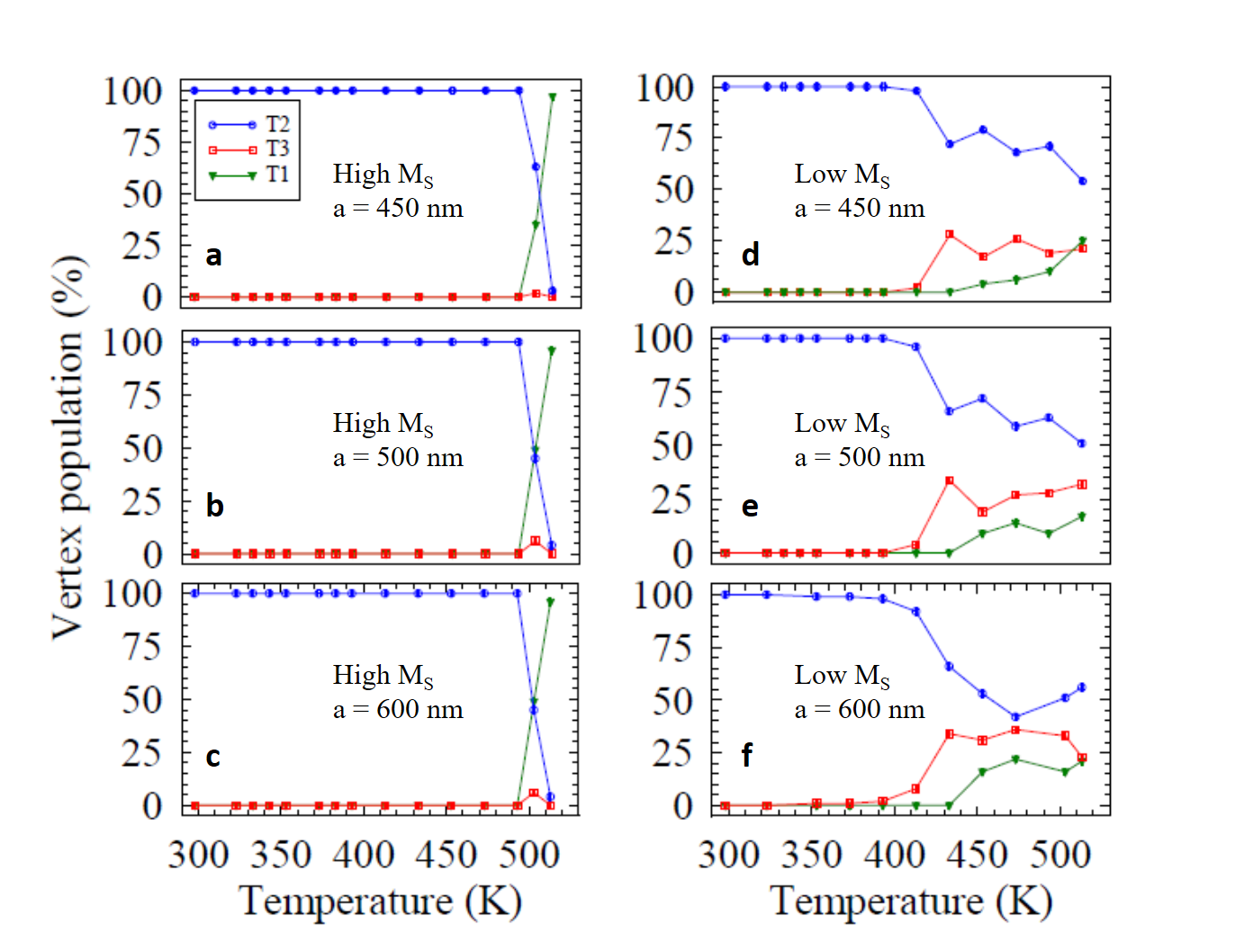}
\caption{The different vertex populations plotted as percentage of the total number of vertices imaged, for the high magnetization sample (left) showing lattice constants (a)~450~nm, (b)~500~nm and (c)~600~nm. The rapid collapse into the highly ordered ground state happens in a small temperature window. The same is plotted for the low magnetization sample (right) with lattice constants (d)~450~nm, (e)~500~nm and (f)~600~nm. This sample has a larger dynamic temperature window but only a low percentage of ground state is observed for all lattice constants.}
\label{fig:vertexevolution}
\end{figure}

Second, we can compare the resulting behavior upon heating between the two samples. The low magnetization sample started to flip its magnetic moments at lower temperatures, after which the vertices gradually changed to a mixture of mainly T$_{1}$, T$_{2}$, and T$_{3}$ vertices. The high magnetization sample required a much higher temperature before any changes started to occur. However, the vertices then rapidly changed to form a long-range ground state configuration via monopole-antimonopole dissociation and fast string growth. Interestingly, for the strongest interacting arrays in the low magnetization sample there is also evidence of monopole-antimonopole dissociation via strings but the vertices created in their path are of the same energy as the initial vertices, it is T$_{2}$ $\rightarrow$ T$_{2}$ conversion. In the much more strongly interacting arrays of the higher magnetization sample the strings follow diagonal paths which create a path of lower energy vertices and it is a T$_{2}$ $\rightarrow$ T$_{1}$ conversion. This demonstrates the increased energy saving to the system as the energy separation between T$_{1}$ and T$_{2}$ vertex types becomes significant enough to change the direction of motion of the magnetic charges. We have used numerical micromagnetic simulations, carried out using the OOMMF code \cite{oommf}, in order to calculate these energies; the results are plotted in Fig~\ref{fig:vertexenergy}. The energy separation between a T$_{1}$ and T$_{2}$ vertex increases as the lattice spacing becomes smaller for both samples. Also as you increase the magnetization you shift the energy scale by an order of magnitude, as you can see from the larger energy scale in Fig~\ref{fig:vertexenergy}a compared to Fig~\ref{fig:vertexenergy}b. The value of the 600~nm high magnetization sample and the 450~nm low magnetization sample are very similar as shown in Fig~\ref{fig:vertexenergy}c, which might suggest the dynamics should also be quite similar once a thermally active temperature has been reached. However, as can be seen from the data the dissociation and the final level of order are very different. This could be due to the M$_{\mathrm{S}}$ values used from the thin film data actually being different to that of the final island material as it has been shown that ion beam milling can affect the magnetic properties of some materials and effectively lower their magnetization \cite{Read2014,Ohsawa2010,McMorran2010}. This highlights the importance of experimental and material considerations for potential device applications where one might want to manipulate the direction of motion of these emergent magnetic charges.

We can plot the evolution of the vertex populations, $P$(T$_{n}$), as a function of temperature in both samples and all lattice spacings, as shown in Fig~\ref{fig:vertexevolution}. It is easy to identify the rapid collapse into the GS across the small temperature window for the high magnetization sample, indicating strong interaction in all three lattice spacings. For the lower magnetization sample we can clearly see the same high level of ground state is not achieved. We can compare this to what would be expected in a random sample. For example, if the nanomagnets were non-interacting and in a completely random arrangement, we would expect the populations, $P$(T$_{1}$), $P$(T$_{4}$) =  12.5~\%. In fact, we observe an excess of ground state vertices, with a maximum of $P$(T$_{1}$)~=~25$\%$ for $a$~=~450~nm, and observed no T$_{4}$ vertices for any of the lattice constants, except for a single observation in the $a$~=~500~nm sample. This shows a degree of interaction is maintained even in the low magnetization sample.


In summary, we have tuned the magnetization strength and Curie temperature of FePd alloys using a simple co-sputtering process. We have fabricated square ASI of different lattice spacings using the lowest and highest magnetization samples. We have imaged their vertex populations at different temperatures using an annealing and subsequent magnetic force microscopy imaging of the arrested states. The samples showed a marked difference in their behavior due to the energy separation of their vertices. In the low magnetization sample we observed an onset of dynamics at much lower temperatures and over a broader range but attained a fairly disordered final state. In contrast, the high magnetization sample showed a rapid collapse towards the ground state over a narrow temperature range. The emergent monopole-antimonopole pairs separated via strings of different excitations. The low magnetization sample tended to separate in vertical chains of T$_{2}$ vertices whereas the high magnetization sample separated via diagonally oriented T$_{1}$ vertices. This study can inform future sample design, in order to study the two different dissociation processes. Fine tuning of the energies can result in drastically different behavior which adds yet another tunable parameter to these statistical mechanical models.

\section*{Methods}
\subsection*{FePd deposition}
The samples were fabricated on SiO$_{2}$ wafer with a 2~nm Pd buffer layer. For the high magnetization sample, Pd and Fe were co-sputtered at powers of 5~W and 10~W, respectively, for a period of 600~s to give an overall FePd thickness of 30~nm. For the low magnetization sample, Pd and Fe were sputtered at powers of 10~W and 13~W, respectively, for a duration of 300~s to give an overall FePd thickness of 33~nm. A Bruker x-ray diffractometer was used to carry out x-ray reflectivity of each sample and also calibration samples. The interference pattern of the scattered x-rays from the interface between different density layers, was fitted to obtain the film thickness using the Kiessig method \cite{Kiessig}. The Fe:Pd ratio was calculated as a ratio of the relative growth rates at the particular powers from calibration sample data and found to be 1:1.9~$\pm$~0.1 for high magnetization sample and 1:3.5~$\pm$~0.1 for the low magnetization sample.
\subsection*{Nanofabrication}
The nanofabrication process for the two ASI samples was identical, and occurred simultaneously. A primer, used to promote resist adhesion, was spin-coated onto the samples at a speed of 4~krpm for 40~s and baked at 180$^{\circ}$C for 60~s. A resist layer of ZEP520A and anisole with a 1:1 ratio was spin-coated onto the samples at 4~krpm for 40 s and baked at 180$^{\circ}$C for 180~s. The ASI pattern was written using e-beam lithography in a JEOL-6300 system with a 100~kV electron beam and a dose of 190~$\mu$C/cm2. The samples were developed for 60~s in N50 solution and rinsed with isopropanol. A 15~nm Ti layer was deposited onto the samples using an e-beam evaporator at a rate of 1~\AA/s. Lift-off was performed using micro-posit remover 1165. Finally, the Ti hard mask layer and unpatterned FePd was removed by Ar ion-beam milling for 70~s, using a 1~kV and 29~mA beam with a 100~V accelerating voltage at a working pressure of $\sim$1$\times$10$^{-5}$~Torr. 
\subsection*{Annealing}
The heating was performed in a rapid thermal annealer where the temperature was controlled using  an infrared tubular halogen lamp furnace and a PID controller which was connected to a thermocouple which was in contact with the carrier wafer. The sample was placed on the carrier wafer and all annealing was done in a N$_{2}$ gas atmosphere to prevent oxidation.
\subsection*{MFM measurements}
Vertex maps were recorded using a Veeco Dimension 3100 magnetic force microscope in tapping mode, using Bruker MESP tips.
\subsection*{Micromagnetic Simulations}
Simulations were carried out using finite element micromagnetic calculations by means of the 3-dimensional Oxsii option of the OOMMF code, in order to calculate the exchange and demagnetizing energies of the different vertex configurations. Rectangular nanoisland shapes with rounded edges were used, discretized into 2$\times$2$\times$30~nm$^{3}$ unit cells. The saturation magnetization was determined from the thin film $M-T$ data in Fig.1; the magnetocrystalline anisotropy constant $\kappa$ was estimated to be zero and the exchange stiffness constant $A$ used was 7.0~pJ/m and 4.4~pJ/m, for high and low $M_{\mathrm{S}}$ respectively. These values were estimated from the percentage of Fe in the alloy. The Gilbert damping coefficient $\alpha$~=~0.5 was used, allowing for rapid convergence (convergence criteria: dm/dt~$\leq$~0.1~deg/ns), after obtaining similar vertex energies using 0.5 and 0.016, where the latter had a much longer simulation time.

\bibliography{MorleyThesis_bib}

\section*{Acknowledgements}
This work was supported by the EPSRC, grant numbers EP/L00285X/1 and EP/L003090/1.

\section*{Contributions}
S.A.M. and M.C.R fabricated the samples. S.A.M. designed the experiment. S.T.R. carried out the heating and microscopy measurements. S.A.M. carried out the magnetometry. S.A.M. and S.T.R. analyzed the results. J.M.P. carried out the OOMMF simulations. E.H.L. provided support for M.C.R.. J.E.C. S.L. and C.H.M. supervised the project. All authors contributed to and reviewed the manuscript.

\section*{Competing Interests}
The authors declare no competing financial interests.

\end{document}